\documentclass[conference]{IEEEtran}
\IEEEoverridecommandlockouts

\usepackage{url}        
\usepackage[colorlinks=true, linkcolor=blue, citecolor=blue, urlcolor=blue]{hyperref}

\usepackage{soul}
\usepackage{cite}
\usepackage{amsmath,amssymb,amsfonts}
\usepackage{algorithmic}
\usepackage{graphicx}
\usepackage{textcomp}
\usepackage{xcolor}
\def\BibTeX{{\rm B\kern-.05em{\sc i\kern-.025em b}\kern-.08em
    T\kern-.1667em\lower.7ex\hbox{E}\kern-.125emX}}

\begin{document}
\title{An Open-Source Python Framework and Synthetic ECG Image Datasets for Digitization, Lead and Lead Name Detection, and Overlapping Signal Segmentation}

\author{
\IEEEauthorblockN{Masoud Rahimi\textsuperscript{1$\dagger$}, Reza Karbasi\textsuperscript{2$\dagger$}, Abdol-Hossein Vahabie\textsuperscript{3}}\\
\IEEEauthorblockA{School of Electrical and Computer Engineering, University of Tehran, Tehran, Iran}\\
\IEEEauthorblockA{\textsuperscript{1}mr.rahimi39@ut.ac.ir, 
\textsuperscript{2}rezakarbasi@ut.ac.ir, 
\textsuperscript{3}h.vahabie@ut.ac.ir}
\thanks{\textsuperscript{$\dagger$}These authors contributed equally to this work.}
}

\date{May 2025}

\maketitle

\begin{abstract}
We introduce an open-source Python framework for generating synthetic ECG image datasets to advance critical deep learning-based tasks in ECG analysis, including ECG digitization, lead region and lead name detection, and pixel-level waveform segmentation. Using the PTB-XL signal dataset, our proposed framework produces four open-access datasets: (1) ECG images in various lead configurations paired with time-series signals for ECG digitization, (2) ECG images annotated with YOLO-format bounding boxes for detection of lead region and lead name, (3)-(4) cropped single-lead images with segmentation masks compatible with U-Net–based models in normal and overlapping versions. In the overlapping case, waveforms from neighboring leads are superimposed onto the target lead image, while the segmentation masks remain clean. The open-source Python framework and datasets are publicly available at \url{https://github.com/rezakarbasi/ecg-image-and-signal-dataset} and \url{https://doi.org/10.5281/zenodo.15484519}, respectively.

\end{abstract}


\section{Introduction}

Electrocardiography (ECG) is a foundational tool in the diagnosis and monitoring of cardiovascular diseases, which remain a leading cause of death worldwide \cite{birnbaum2014role}. Access to ECG time-series data significantly improves the performance of deep learning–based clinical analysis \cite{adedinsewo2022digitizing}. For decades, healthcare institutions have stored ECG records in paper or scanned image formats. These legacy records contain valuable clinical information, including patient history and rare cardiac events \cite{reyna2024ecg}. In many hospitals, ECGs are often stored as images or PDFs rather than raw signals, as this method lowers costs and eliminates the need for specialized hardware or trained personnel \cite{lence2023automatic}. As a result, digitizing ECG images into structured time-series data has become an essential task.

Lence et al. \cite{lence2023automatic} highlighted the lack of open-access ECG digitization datasets for benchmarking. Since then, three large-scale, open-access datasets have been introduced that provide both ECG images and their corresponding ground truth signals: ECG-Image-Kit \cite{shivashankara2024ecg}, ECG-Image-Database \cite{reyna2024ecg}, and PMcardio ECG Image Database \cite{iring2024PMcardio}. All three datasets consist of synthetic ECG images generated from the original waveforms, and they incorporate different augmentations to simulate real-world artifacts. Section~\ref{sec:relatedworks} provides a detailed overview of these datasets.

One of the most critical components of fully automated ECG digitization is the detection of regions of interest, such as lead region and lead names \cite{IA2023ECG, enetcom2023Ecg, ECGArtivatic2021ECG}. Several datasets available on Roboflow include bounding boxes around ECG leads, with a maximum sample size of 1,227. However, no large-scale dataset currently provides bounding boxes for both ECG leads and their corresponding lead names. Furthermore, U-Net \cite{ronneberger2015u} segmentation model has proven effective for extracting waveform traces from individual ECG lead images \cite{li2020deep, demolder2025high}, yet there is currently no public dataset that provides ECG images alongside both segmentation masks and ground truth signals. In addition, overlapping waveforms—a common occurrence in paper ECGs—pose a significant challenge for digitization. A recent Nature article reported that digitization accuracy—measured by correlation with ground truth—dropped below 60\% in the presence of overlapping leads \cite{wu2022fully}. Despite the importance of this problem, no open-access dataset currently includes ECG images with overlapping lead signals paired with clean (non-overlapping) masks. Such a resource could play a key role in training deep learning models to handle this challenge. 

To overcome these limitations in data availability, we present four ECG image datasets from time-series signals. Our proposed datasets support various machine learning tasks, including ECG digitization, waveform segmentation, lead detection, and lead name (e.g., Lead II, V1, etc.) detection and recognition. We release an open-source Python framework to customize and generate large-scale datasets. The datasets include standard lead configurations such as 3$\times$1, 3$\times$4, 6$\times$2, and 12$\times$1, where each format refers to the layout of leads across rows and columns (e.g., 3 rows$\times$4 columns in the 3$\times$4 configuration). We used the PTB-XL signal dataset to generate synthetic ECG images. The dataset for signal segmentation is provided in two versions: normal and overlapping. In the overlapping version, signals from adjacent leads (above or below) are superimposed onto a target lead. At the same time, the corresponding masks remain clean, containing only the true waveform of the target lead. This design enables the training and evaluation of digitization models under more realistic and challenging conditions. The full release includes:

\begin{itemize}
\item An open-source Python framework for generating customizable and large-scale datasets.
\item ECG images paired with time-series signals.
\item ECG images annotated with object detection labels, including bounding boxes in YOLO format around each lead region and its name (e.g., Lead II, V1, etc.).
\item Cropped single-lead images with pixel-level segmentation masks for U-Net-style segmentation (normal and overlapping versions), and paired time-series signals.

\end{itemize}

\section{Related Works}
\label{sec:relatedworks}

Four publicly available datasets have recently emerged as valuable resources for evaluating ECG digitization methods."

Four publicly available datasets, including three large-scale ones, have recently contributed valuable resources for evaluating ECG digitization methods. ECG-Image-Kit \cite{shivashankara2024ecg} is an open-source toolkit for generating synthetic multi-lead ECG images from time-series data, incorporating realistic distortions including text artifacts, wrinkles, and creases on standard ECG paper backgrounds. Using the PhysioNet QT database \cite{laguna1997database}, the authors created a dataset of 21,801 ECG images with corresponding signals. The ECG-Image-Database \cite{reyna2024ecg} extended ECG-Image-Kit by combining its programmatic distortions with real-world physical effects, such as soaking, staining, and mold growth applied to printed ECGs. The physically altered ECGs were scanned or photographed under diverse lighting conditions to capture realistic imaging artifacts. The authors used 977 ECG records from the PTB-XL database and 1,000 from Emory Healthcare to generate a dataset of 35,595 samples. 

PTB-Image \cite{Nguyen2025PTBImageAS} used ECG-Image-Kit to generate 549 ECG images with corresponding signals, based on the PTB signal dataset \cite{bousseljot1995nutzung}, providing paired ECG signals and their corresponding image formats. The PMcardio ECG Image Database \cite{iring2024PMcardio} provides 6000 ECG images derived from 100 waveforms in the PTB-XL dataset. The dataset includes images affected by realistic conditions such as paper bends, crumbles, and capture via scanning or different mobile devices from computer screens. It also features a wide range of augmentation techniques, including contrast and brightness changes, perspective transformations, rotations, blurring, JPEG compression, and resolution changes.

As mentioned earlier, lead detection is an important step for reaching fully automated ECG digitization. 
Several ECG lead detection datasets provide images annotated with bounding boxes around individual leads \cite{IA2023ECG, enetcom2023Ecg, ECGArtivatic2021ECG}. These datasets provide bounding box annotations for object detection tasks in multiple widely used formats, including COCO (JSON), Pascal VOC (XML), and YOLO (TXT). However, these datasets do not contain annotations for both lead name and lead, and the number of samples is limited.

\section{Method}
\begin{figure}[!b]
  \centering
  \includegraphics[scale=0.9]{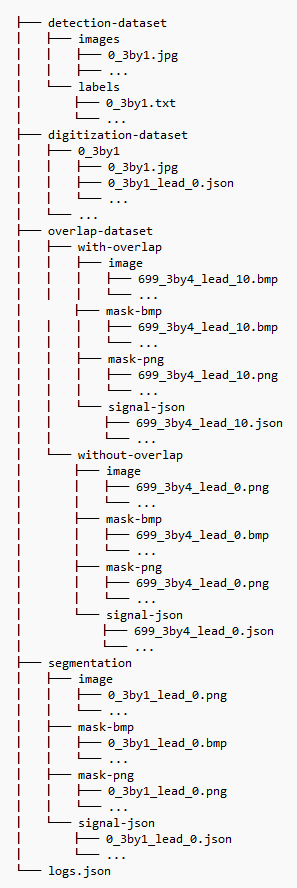}

  \caption{File structure of the ECG dataset release.}
  \label{fig:tree}
\end{figure}

The open-source framework is available at \href{https://github.com/rezakarbasi/ecg-image-and-signal-dataset}{GitHub}, and the open-access datasets \cite{rezakarbasi202515484519} are available at \href{https://doi.org/10.5281/zenodo.15484519}{Zenodo}. The dataset structure is illustrated in Figure~\ref{fig:tree}.

\subsection{Open-source Python Framework and Multiple Datasets}

Our framework allows researchers to generate and customize large-scale ECG image datasets directly from raw time-series signals. Users can easily adjust various parameters, including dataset size (number of samples per split), visual layout (row height, horizontal and vertical scale), file paths for NPY file and output data, and stylistic elements such as grid visibility, lead names, font size, waveform and grid colors, and margin padding. The framework operates on time-series data converted to NumPy format (\texttt{.npy}) from the PTB-XL dataset, which is generated once and reused for efficient processing.

We have released the framework as open-source to support ongoing research and development. Researchers could extend our work by adding simulated real-life artifacts such as text overlays, wrinkles, creases, paper bends, crumples, stains, glare, and scanning distortions; as well as digital augmentations including contrast and brightness adjustments, perspective transformations, rotations, blurring, JPEG compression, resolution changes, and image capture artifacts from mobile devices or computer screens.

\subsection{ECG time-series (PTB-XL) Dataset}
To generate ECG images paired with their corresponding ground truth signals, we used the open-access PTB-XL dataset \cite{wagner2020ptb}. PTB-XL is a large clinical dataset that contains 21,799 12-lead ECG recordings collected from 18,869 patients. Each recording is 10 seconds in duration and includes all standard leads: I, II, III, aVR, aVL, aVF, V1–V6. The patient population is 52\% male and 48\% female, ranging from infants to elderly adults (0–95 years, median age 62). PTB-XL includes a broad spectrum of cardiac conditions, including normal, myocardial infarction, conduction disturbances, hypertrophy, and ST/T. The ECG signals are provided in the WFDB format at a sampling frequency of 500 Hz, with additional 100 Hz downsampled versions. Rich metadata is included in accompanying CSV files, covering diagnostic labels, signal quality indicators, demographic attributes, and structured annotations \cite{wagner2020ptb, reyna2024ecg}.

\subsection{Digitization Dataset}

\begin{figure}[!t]
  \centering
  \includegraphics[width=\columnwidth]{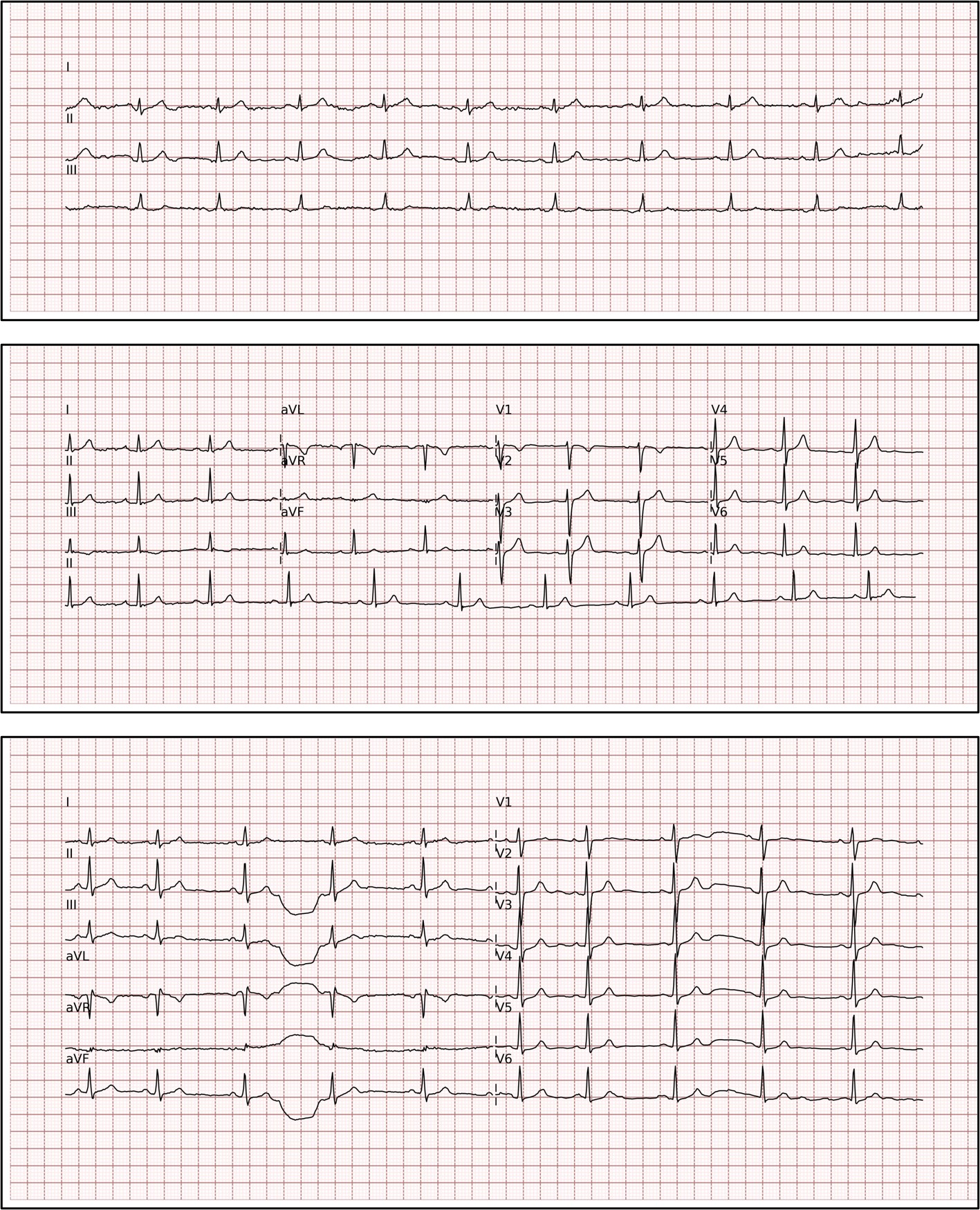}
  \caption{Sample of ECG images in various lead layout configurations.}
  \label{fig:ecg_layouts}
\end{figure}

This dataset contains 2000 synthetic ECG images paired with ground truth signals. It supports multiple lead layouts, including 3$\times$1, 3$\times$4, 6$\times$2, and 12$\times$1. Figure \ref{fig:ecg_layouts} shows ECG images in several lead layout configurations. The ECG image contains separator lines, printed lead names (I, II, III, aVR, aVL, aVF, V1–V6), and the calibrated grid lines. Some layouts, such as 3$\times$4, 6$\times$2, may include lead II repeated in full 10-second length at the bottom of the image. The presence of a full lead, lead name, separating line between two leads, and gridlines can be customized using parameters in the Python framework.

Our pipeline converts raw multi-lead ECG signals directly into \texttt{.jpg} images that resemble standard clinical printouts. Each signal is plotted using Matplotlib on a calibrated grid, with a horizontal scale of {0.2}{s} and a vertical scale of {0.5}{mV}.

\subsection{Lead Detection Dataset}
\begin{figure}[!t]
  \centering
  \includegraphics[width=\columnwidth]{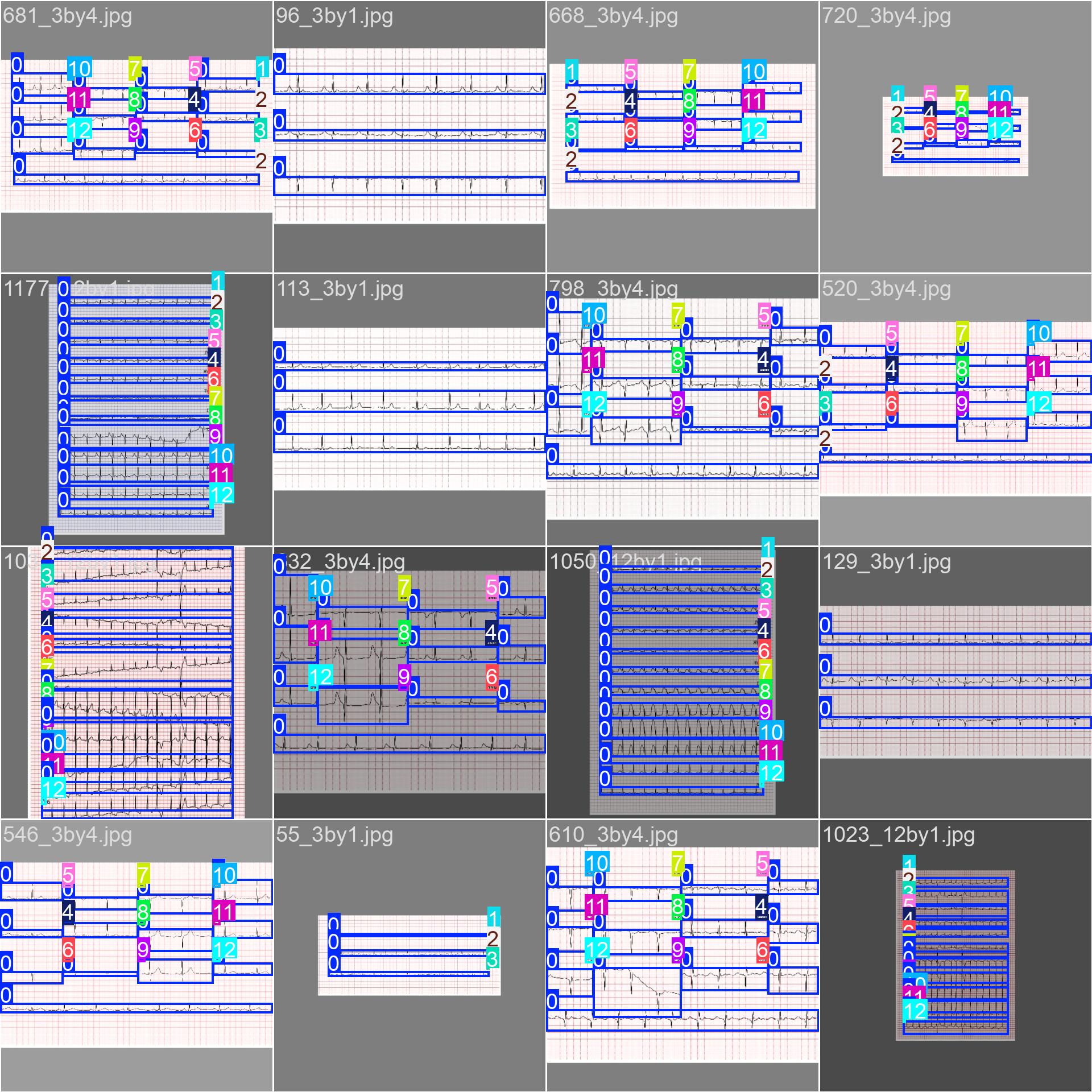}
  \caption{Detection dataset samples: ECG images annotated with bounding boxes for lead regions and lead names. The color of each box indicates the class ID. Class 0 corresponds to lead waveform regions, while classes 1 to 12 represent the lead names I, II, III, aVR, aVL, aVF, and V1–V6, respectively.}
  \label{fig:detection_subplot}
\end{figure}

\begin{figure}[!b]
  \centering
  \includegraphics[width=\columnwidth]{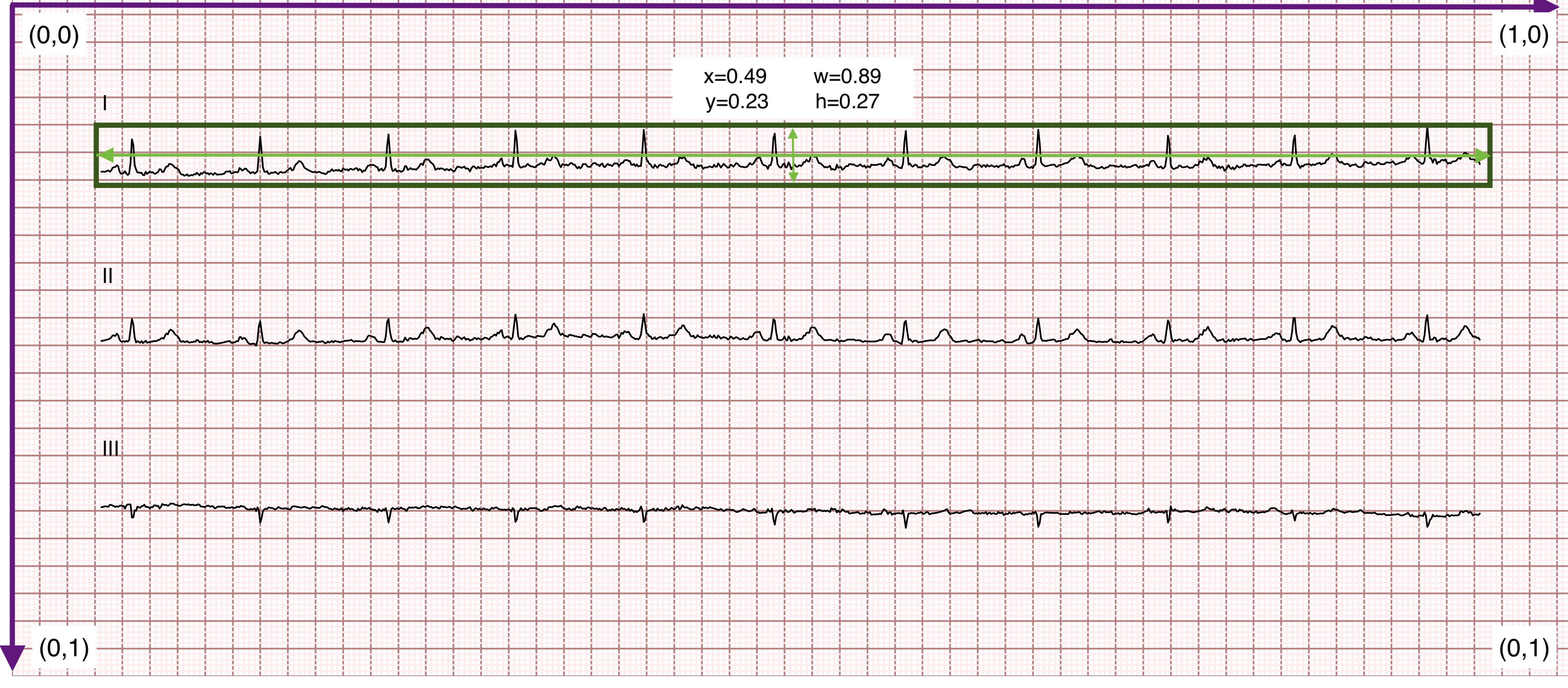}
  \caption{Illustration of the YOLO bounding box format, which includes the class ID, normalized center coordinates $(x, y)$ and the bounding box width and height $(w, h)$, all expressed relative to the image dimensions}
  \label{fig:yolo label description}
\end{figure}

The detection dataset includes 2,000 samples annotated in YOLO format for lead region and lead name detection in ECG images. Figure~\ref{fig:detection_subplot} demonstrates ECG images with different lead layouts and bounding boxes for the lead region and lead name. Each annotation is stored in a \texttt{.txt} file, with one line per object in the image. Each line contains five values: the class index as an integer (\(c\)), the normalized coordinates of the bounding box center \((x, y)\), and the normalized width and height \((w, h)\), all expressed as ratios relative to the image dimensions. The top-left corner of the image is treated as the origin, with the positive \(x\)-axis extending to the right and the positive \(y\)-axis extending downward (see Figure~\ref{fig:yolo label description}). The class index (\(c\)) indicates the type of annotated object: \(c = 0\) denotes lead waveform regions, while \(c = 1\) to \(c = 12\) correspond to the 12 lead names (I, II, III, aVR, aVL, aVF, V1--V6). For example, a bounding box with \(c = 1\) identifies the "Lead I," and \(c = 9\) corresponds to "V3."

\subsection{Normal Segmentation Dataset}
The segmentation dataset includes 20,000 cropped single-lead ECG images, corresponding \texttt{.png} and \texttt{.bmp} masks, and \texttt{.json} time-series files. The top panel of Figure~\ref{fig:segmentation} shows an example of a cropped ECG image alongside its corresponding \texttt{.png} mask. For each cropped lead, two types of masks are generated: a grayscale \texttt{.png} mask and a binary \texttt{.bmp} mask, where foreground pixels are set to 1 and background pixels to 0. The BMP format is particularly useful for training U-Net--style segmentation models.

To create the segmentation dataset, we first cropped each lead region using bounding boxes from the lead detection dataset. For mask generation, ECG images were regenerated in grayscale without grid lines or lead names. The resulting mask images were saved as \texttt{.png} files and then converted to \texttt{.bmp} format to produce binary masks that serve as labels for training U-Net models.

\subsection{Overlapping Segmentation Dataset}
\begin{figure}[!t]
  \centering
  \includegraphics[width=\columnwidth]{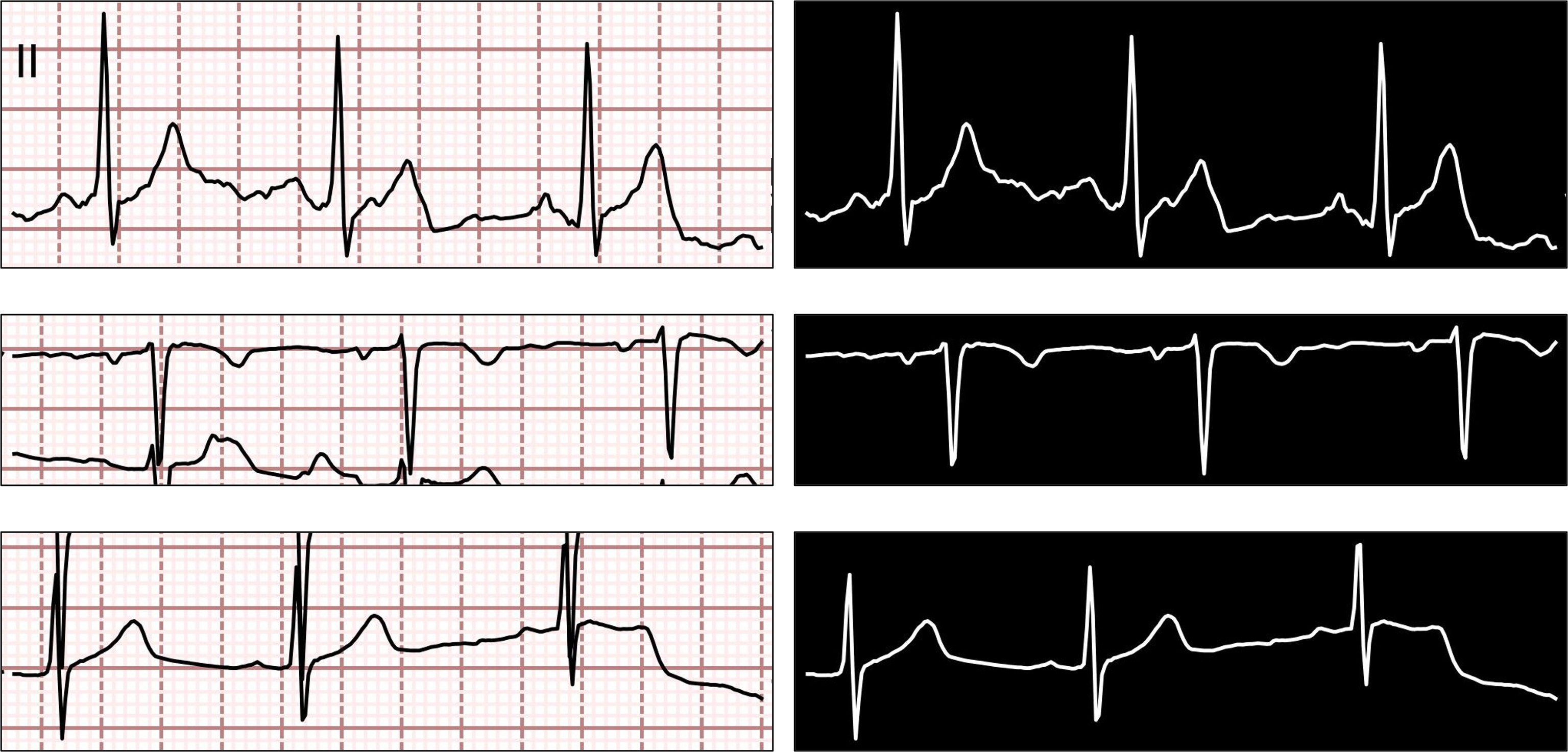}
  \caption{Top—Single-lead ECG image with its corresponding mask shown on the right. Middle and bottom—Overlapping single-lead ECG images with clean masks displayed on the right.}
  \label{fig:segmentation}
\end{figure}

ECG digitization accuracy drops sharply when printed leads overlap—a challenge highlighted in a recent study \cite{wu2022fully}. To address this, we created an overlapping segmentation dataset containing 102 overlapping samples. This dataset contains single-lead ECG images where signals from adjacent leads (above or below) are superimposed onto a target lead. Meanwhile, the segmentation masks remain clean, retaining only the target lead waveform. Examples of overlapping images and their clean masks are shown in the middle and bottom panels of Figure~\ref{fig:segmentation}.

To generate this dataset, we reduced vertical spacing between leads by adjusting the \texttt{row\_height} parameter in our framework. We then reviewed the resulting single-lead images and divided them into two groups: with overlap (102 samples) and without overlap (186 samples). Masks were created by regenerating grayscale images without gridlines and lead names. To maintain clean masks, we only plotted the target lead waveform and omitted the leads in the grayscale image, allowing us to isolate the target lead from overlapping signalsE

\bibliographystyle{ieeetr}
\bibliography{references}
\end{document}